\RequirePackage{fix-cm}

\documentclass[smallcondensed]{svjour3}     
\usepackage[usenames]{color} 
\usepackage{amsmath,amssymb}
\usepackage{graphicx}
\usepackage{mathrsfs}
\usepackage{ulem}
\usepackage{url}
\usepackage{natbib}
\usepackage[linkcolor=blue,citecolor=blue,colorlinks]{hyperref}

\newcommand{\bs}[1]{\ensuremath{\boldsymbol{#1}}}

\journalname{Celest. Mech. Dyn. Astr.}

\begin{document}

\title{The linear stability of the post-Newtonian triangular equilibrium 
in the three-body problem}

\titlerunning{The linear stability of the post-Newtonian triangular equilibrium 
in the three-body problem}

\author{Kei Yamada \and Takuya Tsuchiya}
\institute{K. Yamada \at
Department of Physics, Kyoto University, Kyoto 606-8502, Japan\\
\email{k.yamada@tap.scphys.kyoto-u.ac.jp}
\and
T. Tsuchiya \at
Department of Mathematics, Faculty of Science and Engineering, 
Waseda University, 
Shinjuku, Tokyo 169-8555, Japan
}

\date{Received: date / Accepted: date}

\maketitle

\begin{abstract}
Continuing work initiated in an earlier publication 
[Yamada, Tsuchiya, and Asada, Phys. Rev. D \textbf{91}, 124016 (2015)], 
we reexamine the linear stability of the triangular solution 
in the relativistic three-body problem for general masses 
by the standard linear algebraic analysis.
In this paper, 
we start with the Einstein-Infeld-Hoffman form of equations of motion 
for $N$-body systems in the uniformly rotating frame.
As an extension of the previous work, 
we consider general perturbations to the equilibrium, 
i.e. we take account of perturbations orthogonal to the orbital plane, 
as well as perturbations lying on it.
It is found that 
the orthogonal perturbations depend on each other 
by the first post-Newtonian (1PN) three-body interactions, 
though these are independent of the lying ones 
likewise the Newtonian case.
We also show that 
the orthogonal perturbations do not affect the condition of stability. 
This is because these always precess with two frequency modes; 
the same with the orbital frequency 
and the slightly different one by the 1PN effect.
The same condition of stability with the previous one,
which is valid even for the general perturbations, 
is obtained from the lying perturbations. 
\keywords{Three-body problem \and Triangular equilibrium \and Linear stability \and General relativity \and Post-Newtonian approximation}
\end{abstract}

\section{Introduction}
The direct detections of gravitational waves from merger of binary black hole 
by Advanced LIGO have opened a new window 
to test general relativity \citep{LIGO1,LIGO2,LIGO3,LIGO4}. 
In the near future, 
gravitational waves astronomy will be largely developed 
by a network of gravitational wave detectors 
such as Advanced VIRGO \citep{aVIRGO} and KAGRA \citep{KAGRA}. 
One of the most promising astrophysical sources is 
inspiraling and merging binary compact stars. 
In fact, 
the two events of Advanced LIGO fit well 
with binary black hole mergers \citep{LIGO1,LIGO2,LIGO3,LIGO4}. 

With growing interest, 
gravitational waves involving three-body interactions 
have been discussed 
(e.g., \citep{CIA,GB,Seto1,DSH,MKL}). 
Even in Newtonian gravity, 
the three-body problem is not integrable by analytical methods.
As particular solutions, however, 
Euler and Lagrange found a collinear solution and 
an equilateral triangular one, respectively.
The solutions to the restricted three-body problem, 
where one of the three bodies is a test particle, are known as 
Lagrangian points \citep{Goldstein,Danby,Marchal}. 
In particular, 
Lagrange's equilateral triangular orbit
has stimulated renewed interest 
for relativistic astrophysics 
\citep{THA,Asada,SM,Schnittman,IYA,YA3,YTA,BDEDSG,YA4}. 
Recently, 
a relativistic hierarchical triple system has been discovered 
for the first time \citep{Ransom}, 
and dynamics of such systems has also been studied 
by several authors \citep{BLS,MH,Wen,Thompson,YA-MN,Seto2}.

For three finite masses, in the first post-Newtonian (1PN) approximation,
the existence and uniqueness 
of a post-Newtonian (PN) collinear solution corresponding to Euler's one 
have been shown by Yamada and Asada \citep{YA1,YA2}.
Also, 
Ichita {\it et al.}, including one of the present authors, have shown that 
an equilateral triangular solution is possible at the 1PN order,
if and only if all the three masses are equal \citep{IYA}. 
Generalizing this earlier work, 
Yamada and Asada have found 
a {\it PN triangular} equilibrium solution for general masses
with 1PN corrections to each side length \citep{YA3}.
This PN triangular configuration for general masses is 
not always equilateral
and it recovers the previous results for the restricted three-body case
\citep{Krefetz,Maindl}. 

In Newtonian gravity, 
Gascheau proved that 
Lagrange's equilateral triangular configuration for circular motion is stable 
\citep{Gascheau}, if
\begin{align}
\frac{m_1 m_2 + m_2 m_3 + m_3 m_1}{M^2} < \frac{1}{27} ,
\label{Eq:CoSN}
\end{align}
where $M$ is the total mass.
Routh extended the result to a general law of gravitation $\propto 1/r^k$, 
and found the condition for stability as \citep{Routh}
\begin{align}
\frac{m_1 m_2 + m_2 m_3 + m_3 m_1}{M^2} < 
\frac13 \left( \frac{3 - k}{1 + k} \right)^2 .
\end{align}
The condition of stability \eqref{Eq:CoSN} has recently been corrected 
in the 1PN approximation as \citep{YTA} 
\begin{align}
\frac{m_1 m_2 + m_2 m_3 + m_3 m_1}{M^2} 
+ \frac{15}{2} \frac{m_1 m_2 m_3}{M^3}  \varepsilon < 
\frac{1}{27} \left( 1  - \frac{391}{54} \varepsilon \right) ,
\label{Eq:CoS1PN}
\end{align}
where we define 
\begin{align}
\varepsilon \equiv \left( \frac{G M \omega}{c^3} \right)^{2/3} ,
\end{align}
with 
the common orbital frequency $\omega$ of the system. 
To derive the condition \eqref{Eq:CoS1PN}, 
only the perturbations in the orbital plane are taken into account 
in the previous paper \citep{YTA}.
In Newtonian gravity, 
it is reasonable because 
the perturbations orthogonal to the orbital plane always 
oscillate with the orbital frequency. 
However, 
it is not obvious whether this is the case at the 1PN order. 

Therefore, 
the main purpose of the present paper is 
to take account of 
the perturbations orthogonal to the orbital plane, 
as well as those lying on it, 
in order to show that 
the perturbations orthogonal to the orbital plane always 
oscillate even at the 1PN order 
and they do not affect the condition of stability.
We also derive the condition of stability 
from the motion of lying perturbations 
by the standard linear algebraic analysis.

This paper is organized as follows. 
In Sec. \ref{Sec:Tri}, 
we briefly summarize 
the PN triangular equilibrium solution for three finite masses 
in the corotating frame.
The Perturbations orthogonal to the orbital plane are discussed 
in Sec. \ref{Sec:z}. 
In Sec. \ref{Sec:xy}, 
we consider the perturbations lying on the orbital plane 
in order to derive the condition of stability.
Section \ref{Sec:Con} is devoted to the conclusion.

\section{The PN triangular equilibrium solution in the corotating frame}
\label{Sec:Tri}
Following Ref. \citep{YA3}, 
we summarize a derivation of 
PN triangular equilibrium solution for general masses in this section.
In order to take account of the terms at the 1PN order, 
we employ the Einstein-Infeld-Hoffman (EIH) form of 
the equations of motion for $N$-body systems in uniformly rotating frame
(please see Appendix \ref{App:EIH-EoM} for the derivation):
\begin{align}
\frac{d^2 \bs{r}_K}{d t^2} =&
\sum_{A \neq K} \frac{G m_A}{r_{K A}^3} \bs{r}_{A K}
- 2 ( \bs{\Omega} \times \bs{v}_K )
- ( \bs{\Omega} \cdot \bs{r}_K ) \bs{\Omega} 
+ \Omega^2 \bs{r}_K 
\notag\\
&
+ \sum_{A \neq K} \frac{G m_A}{r_{K A}^3}\bs{r}_{A K}
\left[
- 4 \sum_{B \neq K} \frac{G m_B}{c^2 r_{K B}}
- \sum_{C \neq A} \frac{G m_C}{c^2 r_{A C}}
\left(1 + \frac{\bs{r}_{A K} \cdot \bs{r}_{A C}}{2 r_{C A}^2} \right)
\right.
\notag\\
&
+ \left( \frac{\bs{v}_K + ( \bs{\Omega} \times \bs{r}_K )}{c} \right)^2
+ 2 \left( \frac{\bs{v}_A + ( \bs{\Omega} \times \bs{r}_A )}{c} \right)^2
\notag\\
&
- 4 \left( \frac{\bs{v}_K + ( \bs{\Omega} \times \bs{r}_K )}{c} \right) 
\cdot
\left( \frac{\bs{v}_A + ( \bs{\Omega} \times \bs{r}_A )}{c} \right)
\notag\\
&
\left. 
- \frac{3}{2}
\left\{ \left( \frac{\bs{v}_A + ( \bs{\Omega} \times \bs{r}_A )}{c} \right) 
\cdot \bs{x}_{A K} \right\}^2
\right]
\notag\\
&
- \sum_{A \neq K} \frac{G m_A}{c^2 r_{K A}^2}
\left[ \bs{x}_{A K} \cdot
\left( \frac{4 [ \bs{v}_K + ( \bs{\Omega} \times \bs{r}_K )]
- 3 [ \bs{v}_A + ( \bs{\Omega} \times \bs{r}_A )]}{c} \right) \right]
\notag\\
& \times
\left( \frac{ [ \bs{v}_K + ( \bs{\Omega} \times \bs{r}_K )]
- [ \bs{v}_A + ( \bs{\Omega} \times \bs{r}_A )]}{c} \right) 
\notag\\
&
+ \frac{7}{2} \sum_{A \neq K} \sum_{C \neq A}
\frac{G m_A}{c^2 r_{K A}} \frac{G m_C}{r_{A C}^3} \bs{r}_{C A} ,
\label{Eq:EIH-EoM}
\end{align}
where $\bs{A} \times \bs{B}$ and $\bs{A} \cdot \bs{B}$ denote 
the outer product and the inner product of vectors $\bs{A}$ and $\bs{B}$ 
in the Euclidean space, 
$\bs{r}_K$ and $\bs{v}_K$ are 
the position and velocity of each body in the rotating frame, respectively, 
$\bs{\Omega}$ is a uniform angular velocity of the coordinate 
respect to an inertial frame, 
and we define
\begin{align}
\Omega &\equiv |\bs{\Omega}| , \\
\bs{r}_{A K} &\equiv \bs{r}_A - \bs{r}_K , \\
r_{A K} &\equiv |\bs{r}_{A K}| , \\
\bs{x}_{A K} &\equiv \frac{\bs{r}_{A K}}{r_{A K}} .
\end{align}
In the following, 
we assume circular motion of bodies.

\begin{figure}[h]
\begin{center}
  \includegraphics[width=10cm]{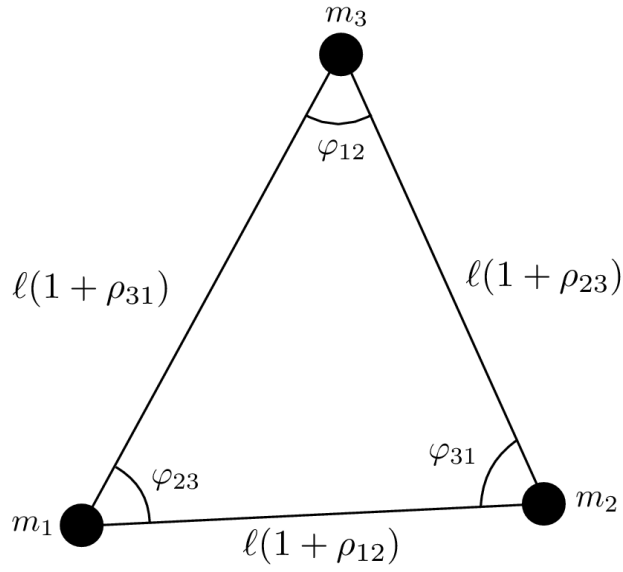}
\caption{
PN triangular configuration.
Each body is located at one of the apexes.
$\rho_{I J}$ denotes the PN corrections to each side length at the 1PN order.
In the equilateral case, $\rho_{12} = \rho_{23} = \rho_{31} = 0$, 
namely, $r_{12} = r_{23} = r_{31} = \ell$ 
according to Eq. \eqref{Eq:mean-length}.}
\label{fig-1}
\end{center}
\end{figure}

Let us consider a PN triangular configuration 
with 1PN corrections to each side length of a Newtonian equilateral triangle, 
so that the distances between the bodies are expressed
\begin{align}
r_{I J} = \ell ( 1 + \rho_{I J} ) , 
\end{align}
where $I, J = 1, 2, 3$ and 
$\rho_{I J} ( = \rho_{J I} )$ are dimensionless PN corrections 
(see Fig. \ref{fig-1}).
Because of circular motion, 
$\ell$ and $\rho_{I J}$ are constants.
Note that we neglect 
the terms of second (and higher) order in $\varepsilon$ henceforth.
Here, if all the three corrections are equal 
(i.e. $\rho_{12} = \rho_{23} = \rho_{31} = \rho$), 
a PN configuration is still an equilateral triangle, 
though each side length is changed by a scale transformation as 
$\ell \to \ell (1 + \rho)$.
Namely, 
one of the degrees of freedom for the PN corrections corresponds to 
a scale transformation, 
and this is unimportant. 
In order to eliminate this degree of freedom, 
we impose a constraint condition 
\begin{align}
\frac{r_{12} + r_{23} + r_{31}}{3} = \ell ,
\label{Eq:mean-length}
\end{align}
which means that 
the arithmetical mean of the three distances of the bodies is not changed 
by the PN corrections.
Namely, 
\begin{align}
\rho_{12} + \rho_{23} + \rho_{31} = 0 .
\end{align}
Please see also Ref. \citep{YA3} for imposing this constraint.

The PN triangular solution for general masses is a coplanar equilibrium, 
in which three bodies rest in Eq. \eqref{Eq:EIH-EoM}, 
therefore
\begin{align}
  \frac{d^2 \bs{r}_K}{d t^2} &= \bs{v}_K = \bs{0},\\
  \bs{\Omega} \cdot \bs{r}_K &= 0,
\end{align}
where we take the origin of the coordinate as the center of mass. 
Straightforward calculations lead to 
\begin{align}
\rho_{12} &= \frac{1}{24} [( \nu_2 - \nu_3 ) ( 5 - 3 \nu_1 ) 
- ( \nu_3 - \nu_1 ) ( 5 - 3 \nu_2 )] \varepsilon , \\
\rho_{23} &= \frac{1}{24} [( \nu_3 - \nu_1 ) ( 5 - 3 \nu_2 ) 
- ( \nu_1 - \nu_2 ) ( 5 - 3 \nu_3 )] \varepsilon , \\
\rho_{31} &= \frac{1}{24} [( \nu_1 - \nu_2 ) ( 5 - 3 \nu_3 ) 
- ( \nu_2 - \nu_3 ) ( 5 - 3 \nu_1 )] \varepsilon , 
\end{align}
with $\nu_I \equiv m_I/M$.
In this case, 
the common orbital frequency is given by 
\begin{align}
\Omega = \omega = \omega_{\rm N} ( 1 + \tilde{\omega}_{\rm PN} ) ,
\label{Eq:ofreq1}
\end{align}
where 
\begin{align}
\omega_{\rm N} &\equiv \sqrt{\frac{G M}{\ell^3}} , 
\label{Eq:ofreq2} \\
\tilde{\omega}_{\rm PN} &\equiv
- \frac{1}{16} ( 29 - 14 V ) \varepsilon ,
\label{Eq:ofreq3}
\end{align} 
with $V \equiv \nu_1 \nu_2 + \nu_2 \nu_3 + \nu_3 \nu_1$.
$V=0$ means two of the three masses are zero, 
thus we consider the case of $V \neq 0$ in this paper.
Hereafter, we take the units of $G = c = 1$.

Before closing this section, 
let us denote perturbations to the positions as 
\begin{align}
\bs{r}_I \to \bs{r}_I + \delta \bs{r}_I ,
\end{align}
and define the relative perturbations as 
\begin{align}
  \delta \bs{r}_{I J} \equiv \delta \bs{r}_I - \delta \bs{r}_J ,
  \label{Eq:relPert}
\end{align}
with 
\begin{align}
\delta \bs{r}_{I J} = 
r_{I J} \left( \xi_{I J} \bs{x}_{I J} + \eta_{I J} \bs{y}_{I J} 
+ \zeta_{I J} \bs{z} \right) ,
\end{align}
where $\bs{z} \equiv \bs{\Omega}/\Omega$ and 
$\bs{y}_{I J} \equiv \bs{z} \times \bs{x}_{I J}$. 
Obviously, we can obtain 
\begin{align}
|\bs{r}_{I J}| \to |\bs{r}_{I J} + \delta \bs{r}_{I J}| 
= r_{I J} ( 1 + \xi_{I J} ) ,
\end{align}
at the 1PN order.
By the definition of $\delta\bs{r}_{I J}$ \eqref{Eq:relPert},
we obtain
\begin{align}
\delta \bs{r}_{12} + \delta \bs{r}_{23} + \delta \bs{r}_{31} = \bs{0} .
\label{Eq:RelpRel}
\end{align}
Note that 
since all degrees of freedom of perturbations are incorporated, 
the orbital energy and angular momentum of the system are 
no longer conservative. 
As we can see below, 
regarding the change of the orbital energy and angular momentum, 
the size and orbital frequency of the triangle, indeed, can be changed.

\section{Orthogonal perturbations} 
\label{Sec:z}
In this section, 
we focus on the perturbations orthogonal to the orbital plane. 
The $z$-direction of perturbed equations of motion is expressed as 
\begin{align}
\ddot{\delta \bs{r}}_I \cdot \bs{z} 
&= 
- \frac{m_J}{\ell^2} \zeta_{I J} + \frac{m_K}{\ell^2} \zeta_{K I} 
\notag\\
&~~~
+ \varepsilon \frac{M}{\ell^2} \left[ 
\frac{1}{24} \nu_J 
\left[ 36 \nu_J^2+36 \nu_J (\nu_K-1)+45 \nu_K^2-18 \nu_K+82\right] \zeta_{I J} 
\right. 
\notag\\
&~~~
-\frac{1}{24} \nu_K 
\left[ 45 \nu_J^2+18 \nu_J (2 \nu_K-1)+36 \nu_K^2-36 \nu_K+82\right] \zeta_{K I} 
\notag\\
&~~~
\left.
-\frac{\sqrt{3}}{2} \nu_J \nu_K 
\left(\frac{\dot{\zeta}_{I J} + \dot{\zeta}_{K I}}{\omega_{\rm N}} \right) 
\right]
,
\label{Eq:pEoM-z}
\end{align}
where the dot denotes the derivatives with respect to time.
Immediately, 
one can find that 
the orthogonal perturbations are independent of the lying ones.
Therefore, 
we can study whether 
a conditions of stability for $z$-direction exists or not, 
separately from the condition for $\xi_{I J}$ and $\eta_{I J}$.

In fact, 
one can infer from Eq. \eqref{Eq:EIH-EoM} that 
the $z$-direction of perturbed equations of motion is separated 
from $\xi_{I J}$ and $\eta_{I J}$.
This is because 
the contributions to the $z$-direction must come 
from the terms parallel to $\bs{\Omega}$ except for $\delta \bs{r}$ itself.
Therefore, 
if $\xi_{I J}$ and $\eta_{I J}$ contribute to the $z$-direction, 
the terms of the form $( \delta \bs{r} \cdot \bs{r} ) \bs{\Omega} $ must appear.
However, in Eq. \eqref{Eq:EIH-EoM}, 
the terms parallel to $\bs{\Omega}$ is only 
of the form $( \bs{\Omega} \cdot \bs{r} ) \bs{\Omega}$.

Moreover, 
we can separate motion of the common center of mass 
from that of the relative perturbations.
The $z$-direction of equations of motion for the common center of mass is 
straightforwardly calculated from Eq. \eqref{Eq:pEoM-z}, 
and this becomes
\begin{align}
\sum_{I = 1}^3 m_I \ddot{\delta \bs{r}}_I \cdot \bs{z} = 0 .
\end{align}
Note that 
although the position of the PN center of mass is different 
from that of the Newtonian one in general, 
we can use the same expressions for 
the orthogonal perturbations in the PN triangular equilibrium.
Therefore, 
the common center of mass is always in uniform linear motion for $z$-direction, 
and then, 
this do not affect the condition of stability.

From Eq. \eqref{Eq:RelpRel}, 
we can find a relation for the relative perturbations of $z$-direction as
\begin{align}
\zeta_{12} + \zeta_{23} + \zeta_{31} = 0 .
\end{align}
Therefore, 
the degrees of freedom of $\zeta_{I J}$ is two. 
Let us eliminate $\zeta_{23}$ by using this relation, 
so that the perturbed equations of motion for $z$-direction can be expressed as
\begin{align}
D \bs{\zeta} &= \mathcal{M} \bs{\zeta} , \label{Eq:pEoM-z2}
\notag\\
\mathcal{M} &\equiv 
\begin{pmatrix}
- 1 + \varepsilon A & \dfrac{1}{2} \nu_3 ( \nu_1 - \nu_2 ) \varepsilon & 
- \dfrac{\sqrt{3}}{2} \nu_2 \nu_3 \varepsilon & 
\dfrac{\sqrt{3}}{2} \nu_3 ( \nu_1 + \nu_2 ) \varepsilon \\
 - \dfrac{1}{2} \nu_2 ( \nu_3 - \nu_1 ) \varepsilon & - 1 + \varepsilon B & 
- \dfrac{\sqrt{3}}{2} \nu_2 (\nu_3 + \nu_1 ) \varepsilon & 
\dfrac{\sqrt{3}}{2} \nu_2 \nu_3 \varepsilon \\
0 & 0 & 1 & 0 \\
0 & 0 & 0 & 1 
\end{pmatrix}
,
\end{align}
where $\bs{\zeta} \equiv (D \zeta_{12}, D \zeta_{31}, \zeta_{12}, \zeta_{31})$ 
with $D \equiv d/ \omega_{\rm N} d t$ 
and 
\begin{align}
A &\equiv 
\frac{1}{8} \left( 6 \nu_1^2+2 \nu_1 \nu_2-6 \nu_1+10 \nu_2^2-10 \nu_2+29 \right) ,
\notag\\
B &\equiv
\frac{1}{8} \left( 6 \nu_1^2+2 \nu_1 \nu_3-6 \nu_1+10 \nu_3^2-10 \nu_3+29 \right) .
\end{align}
The eigenvalues of the matrix $\mathcal{M}$ are 
\begin{align}
    \lambda_1=- i ( 1 + \tilde{\omega}_{\rm PN} ), \quad
    \lambda_2=i ( 1 + \tilde{\omega}_{\rm PN} ), \quad
    \lambda_3=- i ( 1 + \tilde{\omega}_{\rm X} ), \quad
    \lambda_4=i ( 1 + \tilde{\omega}_{\rm X} ) ,
\end{align}
where
\begin{align}
\tilde{\omega}_{\rm X} \equiv
- \frac{1}{16} \left( 29 - 6 V \right) \varepsilon .
\end{align}
Since $\varepsilon \ll 1$, 
each $\lambda$ is always purely imaginary.
Note that 
neglecting the higher order in $\varepsilon$ for the eigenequation leads 
to incorrect eigenvalues, 
because the eigenvalues are degenerate in the Newtonian limit.
Therefore, 
we neglect the higher order after solving the eigenequation.

Since all the eigenvalues are different from each other, 
there is a regular matrix $\mathrm{P}$ such that 
$\mathrm{P}^{-1}\mathcal{M}\mathrm{P} = 
\text{diag}(\lambda_1, \lambda_2, \lambda_3,\lambda_4)$. 
Therefore, 
Eq. \eqref{Eq:pEoM-z2} can be rewritten as 
\begin{align}
    D\hat{\bs{\zeta}} = \mathrm{P}^{-1}\mathcal{M}\mathrm{P}\hat{\bs{\zeta}} ,
\end{align}
where $\hat{\bs{\zeta}}\equiv \mathrm{P}^{-1}\bs{\zeta}$.
We can solve the above equation for $\hat{\bs{\zeta}}$ as 
\begin{align}
    \hat{\bs{\zeta}}
    = \exp(\omega_{\text{N}}t\mathrm{P}^{-1}\mathcal{M}\mathrm{P})
    \hat{\bs{\zeta}}_0
    = \text{diag}(e^{\lambda_1\omega_{\text{N}}t},
    e^{\lambda_2\omega_{\text{N}}t}, e^{\lambda_3\omega_{\text{N}}t},
    e^{\lambda_4\omega_{\text{N}}t})\hat{\bs{\zeta}}_0,
\end{align} 
where $\hat{\bs{\zeta}}_0$ is the initial value. 
Thus, 
the perturbations $\bs{\zeta}$ can be expressed 
by using the trigonometric functions, 
and hence 
$\zeta_{I J}$ always oscillate with two frequency modes;
$( 1 + \tilde{\omega}_{\rm PN} )$ and $( 1 + \tilde{\omega}_{\rm X} )$.
Therefore, the PN triangular equilibrium is stable 
for the perturbations orthogonal to the orbital plane 
likewise the Newtonian case.
It is worthwhile to mention that 
in contrast to the Newtonian case 
$\zeta_{I J}$ has the mode $( 1 + \tilde{\omega}_{\rm X} )$, 
which is different from the orbital frequency.
This might induce {\it resonant orbits} in the nonlinear analysis \citep{MD}.

\section{Lying perturbations}
\label{Sec:xy}

Next, 
we consider the perturbations in the orbital plane.
For these perturbations, 
the motion of the center of mass is slightly complicated.
This is because the PN corrections to the position of the center of mass 
is not canceled for these perturbations.
However, 
the motion of the center of mass is not important 
to consider the stability of the PN triangular configuration.
Hence, 
we focus on the relative perturbations $\xi_{I J}$ and $\eta_{I J}$ 
in this paper.

\begin{figure}[h]
\begin{center}
  \includegraphics[width=10cm]{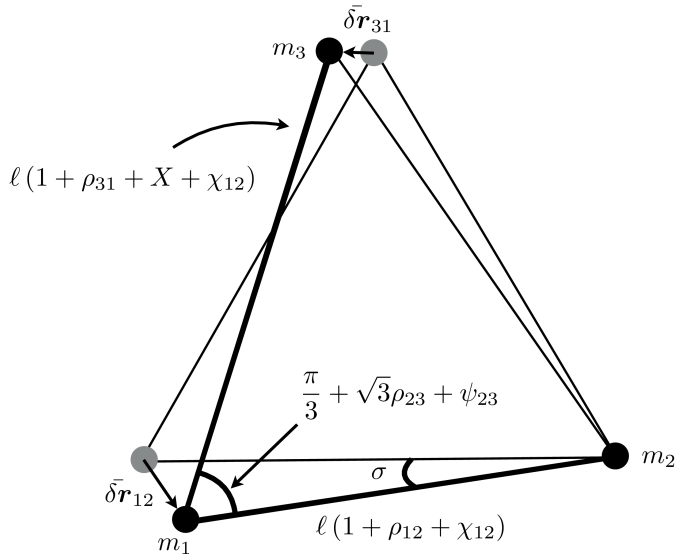}
\caption{
Perturbations in the orbital plane with Routh's variables, 
in which we consider the relative perturbations 
to $\bs{r}_1$ and $\bs{r}_3$ with fixed $\bs{r}_2$.
$\bar{\delta \bs{r}}_{12}$ and $\bar{\delta \bs{r}}_{31}$ 
are the projected vector onto the plane.
$\chi_{12}$ and $\sigma$ correspond to 
the scale transformation of the triangle 
and the change of the angle of the system to a reference direction, 
respectively.
On the other hand, 
$X$ and $\psi_{23}$ are
the degrees of freedom of a shape change
from the equilateral triangle.}
\label{fig-2}
\end{center}
\end{figure}

By separating the motion for the center of mass, 
it is sufficient to discuss the remaining four degrees of freedom 
for $\xi_{I J}$ and $\eta_{I J}$.
For the relative perturbations in the orbital plane, 
it is convenient to use Routh's variables 
$\chi_{12}$, $X$, $\psi_{23}$, and $\sigma$, 
in which we consider the relative perturbations to $\bs{r}_1$ and $\bs{r}_3$
with fixed $\bs{r}_2$.
Here, 
$\chi_{12}$ and $\sigma$ correspond to 
the scale transformation of the triangle 
and the change of the angle of the system to a reference direction, 
respectively.
On the other hand, 
$X$ and $\psi_{23}$ are
the degrees of freedom of a shape change
from the equilateral triangle.
Figure \ref{fig-2} shows the perturbations using Routh's variables. 
By the linear transformations as (please see also Appendix \ref{App:Routh})
\begin{align}
\chi_{12} &= ( 1 + \rho_{12} ) \xi_{12} 
\label{Eq:chi}
, \\
X &= ( 1 + \rho_{31} ) \xi_{31} - ( 1 + \rho_{12} ) \xi_{12} 
\label{Eq:X}
, \\
\psi_{23} &= \eta_{31} - \eta_{12} 
\label{Eq:psi} 
, \\
\sigma &= \eta_{12}
\label{Eq:sigma}
,
\end{align}
we obtain the equations of motion as 
\begin{align}
0 &= 
\left(D^2-3\right) \chi_{12} - 2 D \sigma 
- \frac{9}{4} \nu_3 X - \frac{3}{4} \sqrt{3} \nu_3 \psi_{23}
+ \varepsilon  \left[ 
\left(\frac{1}{32} \left[ -11 \nu_2^2 (9 \nu_3+8) 
\right.
\right.
\right.
\notag\\
&~~~
\left.
+ \nu_2 \left(-72 \nu_3^2 
-34 \nu_3+88\right)+63 \nu_3^3-34 \nu_3^2+16 \nu_3+540\right] 
\notag\\
&~~~
\left.
-\frac{1}{8} \sqrt{3} D \nu_3 (9 \nu_3-7) (2 \nu_2+\nu_3-1)\right) \chi_{12} 
+ \frac{1}{24} D \left[ -6 \nu_2^2 (9 \nu_3+19)
\right.
\notag\\
&~~~
\left.
-6 \nu_2 \left(9 \nu_3^2+10 \nu_3-19\right)+27 \nu_3^3-60 \nu_3^2+63 \nu_3+125\right] \sigma 
\notag\\
&~~~
+ \left(\frac{1}{32} \nu_3 \left[ 99\nu_2^2+2 \nu_2 (27 \nu_3-85)+171 \nu_3^2-304 \nu_3+553\right] 
\right.
\notag\\
&~~~
- \frac{1}{8} \sqrt{3} D \nu_3 \left[ \nu_2 (9 \nu_3+7)
+9 \nu_3^2-12 \nu_3-1\right] \biggr) X 
\notag\\
&~~~
+ \left(\frac{1}{8} D \nu_3 \left[ -\nu_2 (9 \nu_3+7)+9 \nu_3^2-32 \nu_3+11\right] 
\right.
\notag\\
&~~~
+ \frac{1}{32} \sqrt{3} \nu_3 \left[ 24 \nu_2^2 
+\nu_2 (60 \nu_3+62)+87 \nu_3^2-54 \nu_3+122\right] \biggr) \psi_{23} 
\biggr] 
\label{Eq:pEoM1}
,\\
0 &= 
D^2 \sigma + 2 D \chi_{12} 
- \frac{3}{4} \sqrt{3} \nu_3 X 
+\frac{9}{4} \nu_3 \psi_{23} 
+ \varepsilon \left[ 
\left(\frac{1}{8} D \left[ -6 \nu_2^2 (3 \nu_3+5)
\right.
\right.
\right.
\notag\\
&~~~
\left.
-6 \nu_2 \left(3 \nu_3^2+2 \nu_3-5\right)
+9 \nu_3^3-36 \nu_3^2+27 \nu_3-61\right] 
\notag\\
&~~~
\left.
+ \frac{3}{32} \sqrt{3} \nu_3 \left[ -3 \nu_2^2+\nu_2 (24 \nu_3-34)+15 \nu_3^2-26 \nu_3+16\right] \right) \chi_{12} 
\notag\\
&~~~
+ \left(\frac{1}{24} D^2 \left[ 6 \nu_2^2+6 \nu_2 (\nu_3-1)-3 \nu_3^2-12 \nu_3+5\right] 
\right.
\notag\\
&~~~
\left.
+\frac{1}{8} \sqrt{3} D \nu_3 (9 \nu_3-13) (2 \nu_2+\nu_3-1) \right) \sigma 
\notag\\
&~~~
+ \left(\frac{1}{8} D \nu_3 \left(-9 \nu_2 \nu_3+\nu_2+9 \nu_3^2-34 \nu_3+13\right) 
\right.
\notag\\
&~~~
+ \frac{1}{32} \sqrt{3} \nu_3 \left[ 45 \nu_2^2+18 \nu_2 (5 \nu_3-3)+81 \nu_3^2
-72 \nu_3+151\right] \biggr) X 
\notag\\
&~~~
+ \left(\frac{1}{8} \sqrt{3} D \nu_3 \left[ \nu_2 (9 \nu_3-1)+9 \nu_3^2-18 \nu_3+5\right]
\right.
\notag\\
&~~~
 - \frac{9}{32} \nu_3 \left[ 20 \nu_2^2 
+2 \nu_2 (6 \nu_3-7)+13 \nu_3^2-18 \nu_3+50\right] \biggr) \psi_{23} 
\biggr] 
, 
\end{align}
\begin{align}
0 &= 
\left(D^2-3\right) \chi_{12} -2 D \sigma
+\left(D^2+\frac{9 \nu_2}{4}-3\right) X 
+ \left(-2 D-\frac{3 \sqrt{3} \nu_2}{4}\right) \psi_{23} 
\notag\\
&~~~
+ \varepsilon  \left[ 
\left(\frac{1}{8} \sqrt{3} D \nu_2 (9 \nu_2-7) (\nu_2+2 \nu_3-1) 
+ \frac{1}{32} \left[ 63 \nu_2^3-2 \nu_2^2 (36 \nu_3+17)
\right.
\right. 
\right. 
\notag\\
&~~~
\left. 
\left.
+\nu_2 \left(-99 \nu_3^2-34 \nu_3+16\right)-88 \nu_3^2+88 \nu_3+540\right] \right) \chi_{12} 
+ \frac{1}{24} D \left[ 27 \nu_2^3
\right.
\notag\\
&~~~
\left.
-6 \nu_2^2 (9 \nu_3+10)+\nu_2 \left(-54 \nu_3^2-60 \nu_3+63\right)-114 \nu_3^2+114 \nu_3+125\right] \sigma 
\notag\\
&~~~
+ \left(\frac{1}{8} \sqrt{3} D \nu_2 [ \nu_2 (9 \nu_3-4)-21 \nu_3+8 ] 
+ \frac{1}{32} \left[ -108 \nu_2^3-18 \nu_2^2 (7 \nu_3-15)
\right. 
\right. 
\notag\\
&~~~
\left. 
\left. 
+\nu_2 \left(-198 \nu_3^2+136 \nu_3-537\right)-88 \nu_3^2+88 \nu_3+540\right] \right) X 
\notag\\
&~~~
+ \left(\frac{1}{24} D \left[ -9 \nu_2^2 (3 \nu_3-4)+\nu_2 \left(-54 \nu_3^2-39 \nu_3+30\right)-114 \nu_3^2+114 \nu_3+125\right] 
\right. 
\notag\\
&~~~
\left.
\left. 
+ \frac{1}{32} \sqrt{3} \nu_2 \left[ 87 \nu_2^2+6 \nu_2 (10 \nu_3-9)+24 \nu_3^2+62 \nu_3+122\right] \right) \psi_{23} 
\right]
, \\
0 &= 
2 D \chi_{12} + D^2 \sigma 
+ \left(2 D-\frac{3 \sqrt{3} \nu_2}{4}\right) X 
+ \left(D^2-\frac{9 \nu_2}{4}\right) \psi_{23} 
\notag\\
&~~~
+ \varepsilon  \left[
\left(\frac{1}{8} D \left[ 9 \nu_2^3-18 \nu_2^2 (\nu_3+2)-3 \nu_2 \left(6 \nu_3^2+4 \nu_3-9\right)-30 \nu_3^2+30 \nu_3-61\right] 
\right.
\right.
\notag\\
&~~~
\left. 
- \frac{3}{32} \sqrt{3} \nu_2 \left[ 15 \nu_2^2+\nu_2 (24 \nu_3-26)-3 \nu_3^2-34 \nu_3+16\right] \right) \chi_{12} 
\notag\\
&~~~
+ \left(\frac{1}{24} D^2 [ \nu_1 (3 \nu_2-6 \nu_3+5)+\nu_2 (3 \nu_3-10)+5 \nu_3]
\right.
\notag\\
&~~~
\left.
 - \frac{1}{8} \sqrt{3} D \nu_2 (9 \nu_2-13) (\nu_2+2 \nu_3-1)\right) \sigma 
\notag\\
&~~~
+ \left(\frac{1}{8} D \left[ -(9 \nu_3+2) \nu_2^2 +\nu_2 \left(-18 \nu_3^2-13 \nu_3+14\right)-30 \nu_3^2+30 \nu_3-61\right] 
\right.
\notag\\
&~~~
\left.
+ \frac{1}{32} \sqrt{3} \nu_2 \left[ 36 \nu_2^2+6 \nu_2 (3 \nu_3+1)+54 \nu_3^2+48 \nu_3+103\right] \right) X 
\notag\\
&~~~
+ \left(\frac{1}{24} D^2 [ \nu_1 (3 \nu_2-6 \nu_3+5)+\nu_2 (3 \nu_3-10)+5 \nu_3]
\right. 
\notag\\
&~~~
 + \frac{1}{8} \sqrt{3} D \nu_2 [ \nu_2 (4-9 \nu_3)+25 \nu_3-8 ] 
\notag\\
&~~~
\left.
\left.
+ \frac{9}{32} \nu_2 \left[ 13 \nu_2^2+6 \nu_2 (2 \nu_3-3)+20 \nu_3^2-14 \nu_3+50\right] \right) \psi_{23} 
\right]
\label{Eq:pEoM4}
.
\end{align}
These are equivalent to  Eqs. (25)-(28) in Ref. \citep{YTA}.
Note that the above equations do not contain $\sigma$. 
This is consistent with the fact that 
the initial value of $\sigma$ can be zero through 
the appropriate coordinate rotation.

The above equations can be rewritten as
  \begin{align}
    D\bs{\chi}
    = \mathcal{N}\bs{\chi},
    \label{Eq:eigenEqMatrix}
  \end{align}
  where $\bs{\chi}=(D\chi_{12}, D X, D\psi_{23},
 D\sigma, \chi_{12}, X, \psi_{23})$ and the 
 components of the coefficient matrix $\mathcal{N}$ are explicitly written in
 Appendix \ref{App:componentsCM}.

In order to obtain the eigenvalues $\lambda$ of the matrix $\mathcal{N}$, 
let us consider the eigenequation for the matrix $\mathcal{N}$.
This is expressed as 
\begin{align}
  \lambda f(\tau) = 0 ,
\end{align} 
where
\begin{align}
f(\tau)=\tau^3 + C\tau^2 + D\tau + E ,
\label{Eq:EigenEqPN3} 
\end{align}
and we define $\tau=\lambda^2$ and
\begin{align}
  &  C\equiv\frac{1}{4}[8-\varepsilon(77-10V)],\\
  &  D\equiv\frac{1}{16}[4(4+27V)+\varepsilon(378V^2-1265V-162W-308)],\\
  & E\equiv\frac{9}{32}[24V+\varepsilon(126V^2-521V+72W)] ,
  \label{Eq:EigenEqPN2E}
\end{align}
with $W\equiv \nu_1\nu_2\nu_3$.
For the cubic equation $f(\tau)=0$, 
we obtain
\begin{align}
  \Delta &= \frac{-C^2D^2+4C^3E-18C D E+4D^3+27E^2}{27}
  \nonumber\\
  &= \frac{27}{16} (27 V-1) V^2 
  \nonumber\\
  &\quad
+ \frac{30618 V^4-105759 V^3+V^2 (4657-13122 W)+9072 V W-288 W}{64}\varepsilon,
  \label{Eq:discriminant}
\end{align}
where $\Delta$ denotes the discriminant of Eq.\eqref{Eq:EigenEqPN3}.
In general, $f(\tau)=0$ does not have zero root since $E\neq 0$.
On the other hand, if $\tau>0$, $f(\tau)\neq 0$ because of $\varepsilon\ll1$.
Therefore, $f(\tau)=0$ does not have positive real roots.
In addition, if $f(\tau)=0$ has roots of complex numbers, 
the matrix $\mathcal{N}$ has complex eigenvalues which are non-zero real parts.
Thus, if $\Delta <0$, 
all of the roots of $f(\tau)=0$ are negative real numbers.
These can be expressed as \citep{YTA}
\begin{align}
\tau_1 = - 1 + a \varepsilon, \,\,
\tau_{\pm} = \frac{- b \pm \sqrt{b^2 - 4 c}}{2} ,
\label{Eq:root-QE}
\end{align}
with 
\begin{align}
a &= \frac{1}{8 V} ( 77 V - 14 V^2 - 36 W) , \\
b &= 1 - \frac{1}{8 V} ( 77 V - 6 V^2 + 36 W ) \varepsilon , 
\label{coef-b} \\
c &= \frac{27}{4} V - \frac{1}{16} ( 1305 V - 378 V^2 + 162 W) \varepsilon .
\label{coef-c}
\end{align}
One can show that the condition $\Delta<0$ is equivalent to $b^2-4c>0$ 
by straightforward calculations.
It is the necessary 
condition for a stable system
that all of the roots of $f(\tau)=0$ are negative real numbers.
Namely, $b^2-4c>0$ or equivalently at the 1PN order \citep{YTA}
\begin{align}
1- 27 V - \left( \frac{391}{54} + \frac{405}{2} W \right) \varepsilon > 0 .
\label{Eq:CoS1PNb}
\end{align}
This is nothing but Eq. \eqref{Eq:CoS1PN}.

As a result, 
the eigenvalues of the matrix $\mathcal{N}$ are
one zero value and six purely imaginary numbers,
namely
\begin{align}
  \lambda_0&=0,\quad
  \lambda_{1\pm}=\pm i\sqrt{\frac{b-\sqrt{b^2-4c}}{2}},\quad
  \lambda_{2\pm}=\pm i\sqrt{\frac{b+\sqrt{b^2-4c}}{2}},\nonumber\\
  \lambda_{3\pm}&=\pm i\sqrt{1-a\varepsilon}.
\end{align}
One can find that 
there is a regular matrix $\mathrm{Q}$, 
such that $\mathrm{Q}^{-1}\mathcal{N}\mathrm{Q}$ becomes a diagonalized matrix. 
Therefore, Eq. \eqref{Eq:eigenEqMatrix} is rewritten as 
  \begin{align}
    D\bar{\bs{\chi}}
    &= \mathrm{Q}^{-1}\mathcal{N}\mathrm{Q}\bar{\bs{\chi}},
    \label{Eq:EigenEqPN6}\\
    \mathrm{Q}^{-1}\mathcal{N}\mathrm{Q}
    &= \text{diag}(0,\lambda_{1+},\lambda_{1-},\lambda_{2+},
        \lambda_{2-},\lambda_{3+},\lambda_{3-}),
  \end{align}
where $\bar{\bs{\chi}}=\mathrm{Q}^{-1}\bs{\chi}$.
The solution of Eq.\eqref{Eq:EigenEqPN6} can be expressed as
\begin{align}
\bar{\bs{\chi}} &= 
\exp \left( \omega_{\rm N} t \mathrm{Q}^{-1}\mathcal{N}\mathrm{Q} \right) 
\bar{\bs{\chi}}_0 
\notag\\
&= 
\text{diag}(1,e^{\omega_{\rm N} t\lambda_{1+}},e^{\omega_{\rm N} t\lambda_{1-}},
      e^{\omega_{\rm N} t\lambda_{2+}},
      e^{\omega_{\rm N} t\lambda_{2-}},e^{\omega_{\rm N} t\lambda_{3+}},
      e^{\omega_{\rm N} t\lambda_{3-}})
      \bar{\bs{\chi}}_0 , 
  \end{align}
and equivalently, 
\begin{align}
    \bs{\chi}
    &=
    \mathrm{Q}\text{diag}\left(
    1, e^{\omega_{\rm N} t\lambda_{1+}}, e^{\omega_{\rm N}t\lambda_{1-}},
    e^{\omega_{\rm N}t\lambda_{2+}}, e^{\omega_{\rm N}t\lambda_{2-}},
    e^{\omega_{\rm N}t\lambda_{3+}}, e^{\omega_{\rm N}t\lambda_{3-}}
    \right)\mathrm{Q}^{-1}\bs{\chi}_0,
    \label{Eq:solPN}
\end{align}
where $\bar{\bs{\chi}}_0=\mathrm{Q}^{-1}\bs{\chi}_0$ and 
$\bs{\chi}_0$ is the initial value. 
Therefore, we can solve the motion of the perturbations as
    \begin{align}
      \left\{
      \begin{array}{l}
        \chi_{12}
        = C_{11} + C_{12}e^{\omega_{\rm N}t\lambda_{1+}}
        + C_{13}e^{\omega_{\rm N}t\lambda_{1-}}
        + C_{14}e^{\omega_{\rm N}t\lambda_{2+}}
        + C_{15}e^{\omega_{\rm N}t\lambda_{2-}}\\
        \quad\qquad
        + C_{16}e^{\omega_{\rm N}t\lambda_{3+}}
        + C_{17}e^{\omega_{\rm N}t\lambda_{3-}},\\
        X
        = C_{21}
        + C_{22}e^{\omega_{\rm N}t\lambda_{1+}}
        + C_{23}e^{\omega_{\rm N}t\lambda_{1-}}
        + C_{24}e^{\omega_{\rm N}t\lambda_{2+}}
        + C_{25}e^{\omega_{\rm N}t\lambda_{2-}}
        + C_{26}e^{\omega_{\rm N}t\lambda_{3+}}\\
        \qquad
        + C_{27}e^{\omega_{\rm N}t\lambda_{3-}},\\
        \psi_{23}
        = C_{31}
        + C_{32}e^{\omega_{\rm N}t\lambda_{1+}}
        + C_{33}e^{\omega_{\rm N}t\lambda_{1-}}
        + C_{34}e^{\omega_{\rm N}t\lambda_{2+}}
        + C_{35}e^{\omega_{\rm N}t\lambda_{2-}}\\
        \qquad\quad
        + C_{36}e^{\omega_{\rm N}t\lambda_{3+}}
        + C_{37}e^{\omega_{\rm N}t\lambda_{3-}},\\
        \sigma
        = C_{40}
        + C_{41}\omega_{\rm N}t
        + C_{42}e^{\omega_{\rm N}t\lambda_{1+}}
        + C_{43}e^{\omega_{\rm N}t\lambda_{1-}}
        + C_{44}e^{\omega_{\rm N}t\lambda_{2+}}
        + C_{45}e^{\omega_{\rm N}t\lambda_{2-}}
        \\
        \qquad+ C_{46}e^{\omega_{\rm N}t\lambda_{3+}}
        + C_{47}e^{\omega_{\rm N}t\lambda_{3-}},
      \end{array}
      \right.
\label{Eq:sol1PN}
  \end{align}
where $C_{40}$ is an integrate constant and 
the all others of $C_{i j}$ are the constant value determined 
by the initial condition $\bs{\chi}_0$.
Note that 
since the number of the degrees of freedom of $\bs{\chi}_0$ is seven, 
each $C_{i j}$ depends on others through the eigenvectors. 
Especially, 
it is worthwhile to mention that 
$C_{41}$ depends on $C_{11}$ 
consistent with Eqs. \eqref{Eq:ofreq1}-\eqref{Eq:ofreq3} 
(Please see also Appendix \ref{App:eigenv}).
Therefore, 
while $\sigma$ includes the term which is linear in time $t$, 
this term does not affect the change of the shape of the PN triangle, 
but gives only the change of the orbital frequency of the system 
regarding the scale transformation $C_{11}$.
Moreover, 
the solutions $\chi_{12}$, $X$ and $\psi_{23}$ give the oscillation 
around the PN triangular equilibrium since there are only the terms, 
which are the forms as $e^{t\lambda}$ with a purely imaginary number $\lambda$.
Therefore, 
the system is stable 
if and only if the condition \eqref{Eq:CoS1PNb} is satisfied.

\section{Conclusion}
\label{Sec:Con}

We reexamined the linear stability of the PN triangular solution 
for general masses 
by taking account of perturbations orthogonal to the orbital plane, 
as well as perturbations lying on it.

We found that 
the orthogonal perturbations depend on each other 
by the first post-Newtonian (1PN) three-body interactions, 
though these are independent of the lying ones 
likewise the Newtonian case.
We also showed that 
the orthogonal perturbations do not affect the condition of stability. 
This is because these always precess with two frequency modes; 
the same with the orbital frequency 
and the slightly different one, 
which is caused by the 1PN effect and derived for the first time.
The existence of the second frequency mode may induce resonant orbits 
in nonlinear analysis.
This is left as a future work. 
The same condition of stability,
which is valid even for the general perturbations, 
was obtained from the lying perturbations.

\section*{Acknowledgments}

We would like to thank Hideki Asada 
for reading the manuscript. 
K.Y. is grateful to Takahiro Tanaka, Hiroyuki Nakano, 
and Hiroyuki Kitamoto for useful comments and encouragements.
This work was supported in part by JSPS Grant-in-Aid for JSPS Fellows, 
No. 15J01732 (K.Y.).

\appendix
\section{Derivation of EIH equations of motion in uniformly rotating frame}
\label{App:EIH-EoM}

  The EIH equation of motion for $N$-body systems is give by \citep{MTW,LL,Will,AFH}
  \begin{align}
    \frac{d^2\bs{r}_K}{d t^2}
    &= \sum_{A\neq K}\frac{Gm_A}{r_{AK}^3}\bs{r}_{AK}
    \biggl[
      1 - 4\sum_{B\neq K}\frac{Gm_B}{c^2r_{BK}}
      - \sum_{C\neq A}\frac{Gm_C}{c^2r_{CA}}\left(1-\frac{\bs{r}_{AK}\cdot
        \bs{r}_{CA}}{2r^2_{CA}}\right)
      + \frac{v_K^2}{c^2}
      + 2\frac{v_A^2}{c^2}
      \nonumber\\
      &\quad
      - 4\frac{\bs{v}_A\cdot\bs{v}_K}{c^2}
      - \frac{3}{2}\left(\frac{\bs{v}_A}{c}\cdot\bs{x}_{AK}\right)^2
      \biggr]
    - \sum_{A\neq K}\frac{Gm_A}{c^2r^2_{AK}}
    \bs{x}_{AK}\cdot\left(
    3\frac{\bs{v}_A}{c}
    -4\frac{\bs{v}_K}{c}
    \right)\left(\frac{\bs{v}_A}{c} - \frac{\bs{v}_K}{c}\right)
    \nonumber\\
    &\quad
    + \frac{7}{2}\sum_{A\neq K}\sum_{C\neq A}\frac{G^2m_Am_C}{c^2r_{AK}r^3_{CA}}
    \bs{r}_{CA},
  \end{align}
  where $\bs{r}_I$, $\bs{v}_I$, $m_I$ are the position, the velocity,
 and the mass of $I$-th particle, respectively.
  $\bs{r}_{I J}\equiv \bs{r}_I-\bs{r}_J$,
  $r_{I J}\equiv |\bs{r}_{I J}|$, $\bs{x}_{I J}\equiv \bs{r}_{I J}/r_{I J}$, $v_I
  \equiv |\bs{v}_I|$.
  For the above equation, we consider the linear transformation of the function
  $t$.
  In general, a linear transformation from $\mathbb{R}^n$ to $\mathbb{R}^n$
  is given by
  \begin{align}
    \bs{r}'=\mathrm{R}\bs{r},
  \end{align}
  where $\bs{r}, \bs{r}'\in\mathbb{R}^n$, $\mathrm{R}$ is an
  $n\times n$ matrix.
  If $\mathrm{R}$ is a one to one and onto mappings,
  the linear mapping of the first order and second order time derivatives
  of $\bs{r}$ are calculated as
  \begin{align}
    \mathrm{R}\frac{d\bs{r}}{d t}
    &= \frac{d\bs{r}'}{d t} + \mathrm{S}\bs{r}' ,\\
    \mathrm{R}\frac{d^2\bs{r}}{d t^2}
    &= \frac{d^2\bs{r}'}{d t^2} + 2\mathrm{S}\frac{d\bs{r}'}{d t}
    + \mathrm{S}^2\bs{r}' + \frac{d\mathrm{S}}{d t}\bs{r}' ,
  \end{align}
  where
  $\mathrm{S} \equiv -(d\mathrm{R})/(d t)\mathrm{R}^{-1}$.

  If $\mathrm{R}$ is a rotation matrix, $\mathrm{R}$ is the orthogonal
  matrix and the determinant of $\mathrm{R}$ is 1.
  Therefore, the transposition of $\mathrm{R}$ is consistent with the
  inverse of $\mathrm{R}$, namely ${}^t\mathrm{R}=\mathrm{R}^{-1}$,
  and $\mathrm{S}$ becomes skew-symmetric.
  Thus, for three dimensional transformation, we can set $\mathrm{S}$ as
  \begin{align}
    \mathrm{S}=\begin{pmatrix}
    0 & -w_3 & w_2\\
    w_3 & 0 & -w_1\\
    -w_2 & w_1 & 0
    \end{pmatrix},
  \end{align}
  then for all vector $\bs{v}\in\mathbb{R}^{3}$, 
it is satisfied the relation such as $S\bs{v}=\bs{\Omega}\times \bs{v}$, 
where $\bs{\Omega}=(w_1, w_2, w_3)$.

  The EIH equation in a uniformly rotating frame of constant angular velocity
  $\bs{\Omega}$ can be expressed as
  \begin{align}
    \frac{d^2 \bs{r}'_K}{d t^2} =&
    \sum_{A \neq K} \frac{G m_A}{(r'_{K A})^3} \bs{r}'_{A K}
    - 2 ( \bs{\Omega} \times \bs{v}'_K )
    - ( \bs{\Omega} \cdot \bs{r}'_K ) \bs{\Omega}
    + \Omega^2 \bs{r}'_K
    \notag\\
    &
    + \sum_{A \neq K} \frac{G m_A}{(r'_{K A})^3}\bs{r}'_{A K}
    \left[
      - 4 \sum_{B \neq K} \frac{G m_B}{c^2 r'_{K B}}
      - \sum_{C \neq A} \frac{G m_C}{c^2 r'_{A C}}
      \left(1 + \frac{\bs{r}'_{A K} \cdot \bs{r}'_{A C}}{2 (r'_{C A})^2} \right)
      \right.
      \notag\\
      &
      + \left( \frac{\bs{v}'_K + ( \bs{\Omega} \times \bs{r}'_K )}{c} \right)^2
      + 2 \left( \frac{\bs{v}'_A + ( \bs{\Omega} \times \bs{r}'_A )}{c} \right)^2
      \notag\\
      &
      \left. 
      - 4 \left( \frac{\bs{v}'_K + ( \bs{\Omega} \times \bs{r}'_K )}{c} \right) 
      \cdot
      \left( \frac{\bs{v}'_A + ( \bs{\Omega} \times \bs{r}'_A )}{c} \right)
      - \frac{3}{2}
      \left\{ \left( \frac{\bs{v}'_A + ( \bs{\Omega} \times \bs{r}'_A )}{c} \right) 
      \cdot \bs{x}'_{A K} \right\}^2
      \right]
    \notag\\
    &
    - \sum_{A \neq K} \frac{G m_A}{c^2 (r'_{K A})^2}
    \left[ \bs{x}'_{A K} \cdot
      \left( \frac{4 [ \bs{v}'_K + ( \bs{\Omega} \times \bs{r}'_K )]
        - 3 [ \bs{v}'_A + ( \bs{\Omega} \times \bs{r}'_A )]}{c} \right) \right]
    \notag\\
    & \times
    \left( \frac{ [ \bs{v}'_K + ( \bs{\Omega} \times \bs{r}'_K )]
      - [ \bs{v}'_A + ( \bs{\Omega} \times \bs{r}'_A )]}{c} \right) 
    \notag\\
    &
    + \frac{7}{2} \sum_{A \neq K} \sum_{C \neq A}
    \frac{G m_A}{c^2 r'_{K A}} \frac{G m_C}{(r'_{A C})^3} \bs{r}'_{C A},
  \end{align}
  with using the relations such as
  \begin{align}
    \bs{r}_{I J}\cdot\bs{r}_{MN}
    &= \bs{r}'_{I J}\cdot\bs{r}'_{MN},\\
    \bs{v}_I\cdot\bs{v}_J
    &= (\bs{v}'_I + \bs{\Omega}\times\bs{r}'_I)\cdot
    (\bs{v}'_J + \bs{\Omega}\times\bs{r}'_J),\\
    \bs{v}_I\cdot\bs{r}_{J K}
    &= (\bs{r}'_I + \bs{\Omega}\times\bs{r}'_{I})\cdot\bs{r}'_{J K} , 
  \end{align}
  and we set $\Omega\equiv|\bs{\Omega}|$

\section{Transformation to Routh's variables}
\label{App:Routh}

Eliminating $\xi_{23}$ and $\eta_{23}$ from Eq. \eqref{Eq:RelpRel}, 
the perturbed equations of motion for 
$\xi_{12}$, $\eta_{12}$, $\xi_{31}$, $\eta_{31}$ become
\begin{align}
0 &= 
\left(D^2+\frac{9 \nu_3}{4}-3\right) \xi_{12} 
+\left(\frac{3 \sqrt{3} \nu_3}{4}-2 D\right) \eta_{12} -\frac{3}{4} \sqrt{3} \nu_3 \eta_{31} -\frac{9}{4} \nu_3 \xi_{31}  
\notag\\
&~~~
+ \varepsilon  \left[ 
\left(\frac{1}{8} \sqrt{3} D \nu_3 (-9 \nu_2 \nu_3+21 \nu_2+4 \nu_3-8)
\right.
\right.
\notag\\
&~~~
\left.
+ \frac{1}{32} \left[ -22 \nu_2^2 (9 \nu_3+4)+\nu_2 \left(-126 \nu_3^2+136 \nu_3+88\right)-108 \nu_3^3+270 \nu_3^2-537 \nu_3+540\right] \right) \xi_{12} 
\notag\\
&~~~
+ \left(\frac{1}{8} D \left[-2 \nu_2^2 (9 \nu_3+17)+\nu_2 \left(-9 \nu_3^2-9 \nu_3+34\right)+10 \nu_3^2+2 \nu_3+45\right]
\right.
\notag\\
&~~~
\left.
- \frac{1}{32} \sqrt{3} \nu_3 \left[30 \nu_2^2+\nu_2 (66 \nu_3+56)+84 \nu_3^2-66 \nu_3+127\right] \right) \eta_{12} 
\notag\\
&~~~
+ \left(\frac{1}{32} \nu_3 \left[126 \nu_2^2+2 \nu_2 (27 \nu_3-76)+144 \nu_3^2-322 \nu_3+553\right] 
\right.
\notag\\
&~~~
\left.
-\frac{1}{8} \sqrt{3} D \nu_3 \left[\nu_2 (9 \nu_3+7)+9 \nu_3^2-12 \nu_3-1\right] \right) \xi_{31} 
\notag\\
&~~~
+ \left(\frac{1}{8} D \nu_3 \left[-\nu_2 (9 \nu_3+7)+9 \nu_3^2-32 \nu_3+11\right] 
\right.
\notag\\
&~~~
\left.
\left.
+ \frac{1}{32} \sqrt{3} \nu_3 \left[ 30 \nu_2^2+\nu_2 (66 \nu_3+56)+84 \nu_3^2-66 \nu_3+127\right] \right) \eta_{31} 
\right] 
, 
\label{Eq:EoMxi12}\\
0 &= 
\left(D^2-\frac{9 \nu_3}{4}\right) \eta_{12} 
+\left(2 D+\frac{3 \sqrt{3} \nu_3}{4}\right) \xi_{12} 
+ \frac{9}{4} \nu_3 \eta_{31} -\frac{3}{4} \sqrt{3} \nu_3 \xi_{31}  
\notag\\
&~~~
+ \varepsilon \left[ 
\left(\frac{1}{8} D \left[ -6 \nu_2^2 (3 \nu_3+5)+\nu_2 \left(-9 \nu_3^2-13 \nu_3+30\right)-2 \nu_3^2+14 \nu_3-61\right] 
\right.
\right. 
\notag\\
&~~~
\left.
-\frac{1}{32} \sqrt{3} \nu_3 \left[ 54 \nu_2^2+6 \nu_2 (3 \nu_3+8)+36 \nu_3^2+6 \nu_3+103\right] \right) \xi_{12} 
\notag\\
&~~~
+ \left(\frac{1}{8} \sqrt{3} D \nu_3 [ \nu_2 (9 \nu_3-25)-4 \nu_3+8] 
\right.
\notag\\
&~~~
\left.
+\frac{3}{32} \nu_3 \left[ 66 \nu_2^2+6 \nu_2 (7 \nu_3-8)+36 \nu_3^2-66 \nu_3+155\right] \right) \eta_{12} 
\notag\\
&~~~
+ \left(\frac{1}{8} D \nu_3 \left(-9 \nu_2 \nu_3+\nu_2+9 \nu_3^2-34 \nu_3+13\right) 
\right.
\notag\\
&~~~
\left.
+ \frac{1}{32} \sqrt{3} \nu_3 \left[ 54 \nu_2^2+6 \nu_2 (15 \nu_3-8)+72 \nu_3^2-78 \nu_3+151\right] \right) \xi_{31} 
\notag\\
&~~~
+ \left(\frac{1}{8} \sqrt{3} D \nu_3 \left[ \nu_2 (9 \nu_3-1)+9 \nu_3^2-18 \nu_3+5\right] 
\right.
\notag\\
&~~~
\left.
\left.
- \frac{3}{32} \nu_3 \left[ 66 \nu_2^2+6 \nu_2 (7 \nu_3-8)+36 \nu_3^2-66 \nu_3+155\right] \right) \eta_{31} 
\right] 
, 
\end{align}
\begin{align}
0 &= 
\left(D^2+\frac{9 \nu_2}{4}-3\right) \xi_{31} 
+\left(-2 D-\frac{3 \sqrt{3} \nu_2}{4}\right) \eta_{31} 
+ \frac{3}{4} \sqrt{3} \nu_2 \eta_{12} -\frac{9}{4} \nu_2 \xi_{12}  
\notag\\
&~~~
+ \varepsilon \left[ 
\left(\frac{1}{8} \sqrt{3} D \nu_2 \left[ 9 \nu_2^2+3 \nu_2 (3 \nu_3-4)+7 \nu_3-1\right] 
\right.
\right. 
\notag\\
&~~~
\left.
+ \frac{1}{32} \nu_2 \left[ 144 \nu_2^2+\nu_2 (54 \nu_3-322)+126 \nu_3^2-152 \nu_3+553\right] \right) \xi_{12} 
\notag\\
&~~~
+ \left(\frac{1}{8} D \nu_2 \left[ 9 \nu_2^2-\nu_2 (9 \nu_3+32)-7 \nu_3+11\right] 
\right. 
\notag\\
&~~~
\left.
- \frac{1}{32} \sqrt{3} \nu_2 \left[ 84 \nu_2^2+66 \nu_2 (\nu_3-1)+30 \nu_3^2+56 \nu_3+127\right] \right) \eta_{12} 
\notag\\
&~~~
+ \left(\frac{1}{8} \sqrt{3} D \nu_2 [\nu_2 (9 \nu_3-4)-21 \nu_3+8] 
\right. 
\notag\\
&~~~
\left. 
+ \frac{1}{32} \left[ -108 \nu_2^3-18 \nu_2^2 (7 \nu_3-15)+\nu_2 \left(-198 \nu_3^2+136 \nu_3-537\right)-88 \nu_3^2+88 \nu_3+540\right] \right) \xi_{31} 
\notag\\
&~~~
+ \left(\frac{1}{8} D \left[ \nu_2^2 (10-9 \nu_3)+\nu_2 \left(-18 \nu_3^2-9 \nu_3+2\right)-34 \nu_3^2+34 \nu_3+45\right] 
\right. 
\notag\\
&~~~
\left. 
\left. 
+ \frac{1}{32} \sqrt{3} \nu_2 \left[ 84 \nu_2^2+66 \nu_2 (\nu_3-1)+30 \nu_3^2+56 \nu_3+127\right] \right) \eta_{31} 
\right]
, \\
0 &= 
\left(D^2-\frac{9 \nu_2}{4}\right) \eta_{31} 
+\left(2 D-\frac{3 \sqrt{3} \nu_2}{4}\right) \xi_{31} 
+ \frac{9}{4} \nu_2 \eta_{12} +\frac{3}{4} \sqrt{3} \nu_2 \xi_{12}  
\notag\\
&~~~
+ \varepsilon \left[
\left(\frac{1}{8} D \nu_2 \left[ 9 \nu_2^2-\nu_2 (9 \nu_3+34)+\nu_3+13\right] 
\right. 
\right. 
\notag\\
&~~~
\left. 
- \frac{1}{32} \sqrt{3} \nu_2 \left[ 72 \nu_2^2+\nu_2 (90 \nu_3-78)+54 \nu_3^2-48 \nu_3+151\right] \right) \xi_{12} 
\notag\\
&~~~
+ \left(\frac{1}{8} \sqrt{3} D \nu_2 \left[ -9 \nu_2^2-9 \nu_2 (\nu_3-2)+\nu_3-5\right] 
\right. 
\notag\\
&~~~
\left. 
- \frac{3}{32} \nu_2 \left[ 36 \nu_2^2+6 \nu_2 (7 \nu_3-11)+66 \nu_3^2-48 \nu_3+155\right] \right) \eta_{12} 
\notag\\
&~~~
+ \left(\frac{1}{8} D \left[ \nu_2^2 (-(9 \nu_3+2))+\nu_2 \left(-18 \nu_3^2-13 \nu_3+14\right)-30 \nu_3^2+30 \nu_3-61\right] 
\right.
\notag\\
&~~~
\left. 
+ \frac{1}{32} \sqrt{3} \nu_2 \left[ 36 \nu_2^2+6 \nu_2 (3 \nu_3+1)+54 \nu_3^2+48 \nu_3+103\right] \right) \xi_{31} 
\notag\\
&~~~
+ \left(\frac{1}{8} \sqrt{3} D \nu_2 [ \nu_2 (4-9 \nu_3)+25 \nu_3-8 ] 
\right.
\notag\\
&~~~
\left.\left.
+ \frac{3}{32} \nu_2 \left[ 36 \nu_2^2+6 \nu_2 (7 \nu_3-11)+66 \nu_3^2-48 \nu_3+155\right] \right) \eta_{31} 
\right] 
.
\label{Eq:EoMeta31}
\end{align}

Let us seek the relations between $(\xi_{I J}, \eta_{I J})$ 
and Routh's variables.
First, 
since $\chi_{12}$ is a perturbation to $r_{12}$, 
we obtain the relation as
\begin{align}
\ell ( 1 + \rho_{12} ) ( 1 + \xi_{12} ) = \ell ( 1 + \rho_{12} + \chi_{12}) .
\end{align}
Therefore, 
\begin{align}
\chi_{12} = ( 1 + \rho_{12} ) \xi_{12} .
\end{align}
In the same way, we obtain the relation for $X$ as
\begin{align}
X = ( 1 + \rho_{31} ) \xi_{31} - ( 1 + \rho_{12} ) \xi_{12} .
\end{align}
Next, 
let us define the projection of vectors onto the orbital plane 
as 
\begin{align}
\bar{\bs{A}} \equiv \bs{A} - ( \bs{A} \cdot \bs{z} ) \bs{z} .
\end{align}
Using this, 
$\sigma$ is expressed as a perturbation to the angle between 
$\bar{\bs{r}}_{12}$ and $\bar{\bs{r}}_{12} + \bar{\delta \bs{r}}_{12}$, 
and then, 
\begin{align}
\sin \sigma = 
\frac{\left|\bar{\bs{r}}_{12} \times ( \bar{\bs{r}}_{12} + \bar{\delta \bs{r}}_{12} ) \right|}{r_{12}^2 ( 1 + \xi_{12} )} . 
\end{align}
Solving this for $\sigma$ to the 1PN order, 
we obtain
\begin{align}
\sigma = \eta_{12} .
\end{align}
Finally, 
since $\psi_{23}$ is a perturbation in the opposite angle of $r_{23}$, 
we obtain 
\begin{align}
\cos \left( \frac{\pi}{3} + \sqrt{3} \rho_{23} + \psi_{23} \right) = 
- \frac{( \bar{\bs{r}}_{31} + \bar{\delta \bs{r}}_{31} ) \cdot ( \bar{\bs{r}}_{12} + \bar{\delta \bs{r}}_{12} )}{r_{31} r_{12} ( 1 + \xi_{31} ) ( 1 + \xi_{12} )} .
\end{align}
This leads to 
\begin{align}
\psi_{23} = \eta_{31} - \eta_{12} ,
\end{align}
at the 1PN order.
Using these relations, 
we obtain the perturbed equations of motion \eqref{Eq:pEoM1}-\eqref{Eq:pEoM4}.

\section{The components of the coefficient matrix of the perturbation
  equation}
\label{App:componentsCM}

The components of the coefficient matrix $\mathcal{N}$ 
in Eq. \eqref{Eq:eigenEqMatrix} are written as
\begin{align}
  \mathcal{N}_{11}
  &=
  \frac{\varepsilon}{8} \sqrt{3} \nu_3 (9 \nu_3-7) (2 \nu_2+\nu_3-1),\\
  \mathcal{N}_{12}
  &=
  \frac{\varepsilon}{8} \sqrt{3} \nu_3 [\nu_2 (9 \nu_3+7)
    +9 \nu_3^2-12 \nu_3-1],\\
  \mathcal{N}_{13}
  &=
  -\frac{\varepsilon}{8} \nu_3 [-\nu_2 (9 \nu_3+7)+9 \nu_3^2-32 \nu_3+11],\\
  \mathcal{N}_{14}
  &=
  2-\frac{\varepsilon}{24} [ -6 \nu_2^2 (9 \nu_3+19)
    -6 \nu_2 \left(9 \nu_3^2+10 \nu_3-19\right)
    +27 \nu_3^3-60 \nu_3^2+63 \nu_3
    \nonumber\\
    &\quad
    +125],\\
  \mathcal{N}_{15}
  &=
  3 - \frac{\varepsilon}{32}[ -11 \nu_2^2 (9 \nu_3+8)
    + \nu_2 \left(-72 \nu_3^2 -34 \nu_3+88\right)
    +63 \nu_3^3-34 \nu_3^2+16 \nu_3
    \nonumber\\
    &\quad+540],\\
  \mathcal{N}_{16}
  &=
  \frac{9}{4}\nu_3
  - \frac{\varepsilon}{32} \nu_3 [99 \nu_2^2+2 \nu_2 (27 \nu_3-85)
    +171 \nu_3^2-304 \nu_3+553],\\
  \mathcal{N}_{17}
  &=
  \frac{3}{4}\sqrt{3}\nu_3
  - \frac{\varepsilon}{32} \sqrt{3} \nu_3[ 24 \nu_2^2 
    +\nu_2 (60 \nu_3+62)+87 \nu_3^2-54 \nu_3+122],\\
  \mathcal{N}_{21}
  &=
  - \frac{\varepsilon}{8} \sqrt{3}\nu_2 (9 \nu_2-7) (\nu_2+2 \nu_3-1)
  -\frac{\varepsilon}{8} \sqrt{3} \nu_3 (9 \nu_3-7) (2 \nu_2+\nu_3-1),\\
  \mathcal{N}_{22}
  &=
  -\frac{\varepsilon}{8} \sqrt{3}\nu_2 [ \nu_2 (9 \nu_3-4)-21 \nu_3+8 ]
  - \frac{\varepsilon}{8} \sqrt{3} \nu_3 [\nu_2 (9 \nu_3+7)
    +9 \nu_3^2-12 \nu_3-1],\\
  \mathcal{N}_{23}
  &=
  2-
  \frac{\varepsilon}{24}[-9 \nu_2^2 (3 \nu_3-4)+3\nu_2 \left(-18 \nu_3^2-13
    \nu_3+10\right)-114 \nu_3^2+114 \nu_3+125]
  \nonumber\\
  &\quad
  + \frac{\varepsilon}{8} \nu_3 [-\nu_2 (9 \nu_3+7)+9 \nu_3^2-32 \nu_3+11],\\
  \mathcal{N}_{24}
  &=
  - \frac{\varepsilon}{24}[27 \nu_2^3-6 \nu_2^2 (9 \nu_3+10)+3\nu_2
    \left(-18 \nu_3^2-20\nu_3+21\right)-114 \nu_3^2+114 \nu_3
    \nonumber\\
    &\quad
    +125]
  +
  \frac{\varepsilon}{24} [ -6 \nu_2^2 (9 \nu_3+19)
    -6 \nu_2 \left(9 \nu_3^2+10 \nu_3-19\right)
    +27 \nu_3^3-60 \nu_3^2
    \nonumber\\
    &\quad
    +63 \nu_3+125]
  ,\\
  \mathcal{N}_{25}
  &=
  - \frac{\varepsilon}{32}[63 \nu_2^3-2 \nu_2^2 (36 \nu_3+17)+\nu_2
    \left(-99 \nu_3^2-34 \nu_3+16\right)-88 \nu_3^2+88 \nu_3
    \nonumber\\
    &\quad
    +540]
  + \frac{\varepsilon}{32}
  [ -11 \nu_2^2 (9 \nu_3+8) 
    + \nu_2 \left(-72 \nu_3^2 
    -34 \nu_3+88\right)+63 \nu_3^3-34 \nu_3^2
    \nonumber\\
    &\quad
    +16 \nu_3+540],
\end{align}
\begin{align}
  \mathcal{N}_{26}
  &=
  -\frac{9\nu_2}{4}+3
  - \frac{\varepsilon}{32}[ -108 \nu_2^3-18 \nu_2^2 (7 \nu_3-15)+\nu_2
    \left(-198 \nu_3^2+136 \nu_3-537\right)
    \nonumber\\
    &\quad-88 \nu_3^2+88 \nu_3+540]
  - \frac{9}{4}\nu_3
  + \frac{\varepsilon}{32} \nu_3 [99 \nu_2^2+2 \nu_2 (27 \nu_3-85)
    +171 \nu_3^2
    \nonumber\\
    &\quad
    -304 \nu_3+553],\\
  \mathcal{N}_{27}
  &=
  \frac{3\sqrt{3}\nu_2}{4}
  - \frac{\varepsilon}{32} \sqrt{3} \nu_2 [87 \nu_2^2+6 \nu_2 (10 \nu_3-9)
    +24 \nu_3^2+62 \nu_3+122]
  - \frac{3}{4}\sqrt{3}\nu_3
  \nonumber\\
  &\quad
  + \frac{\varepsilon}{32} \sqrt{3} \nu_3[ 24 \nu_2^2 
    +\nu_2 (60 \nu_3+62)+87 \nu_3^2-54 \nu_3+122]
  ,\\
  \mathcal{N}_{31}
  &=
  - \frac{\varepsilon}{24} [ 27 \nu_2^3-54 \nu_2^2 (\nu_3+2)
    -\nu_2 \left(54\nu_3^2+42\nu_3-101\right)-90 \nu_3^2+80 \nu_3-183
    \nonumber\\
    &\quad
    -2\nu_1(3\nu_2-6\nu_3+5)]
  +
  \frac{\varepsilon}{24}[ -6 \nu_2^2 (9 \nu_3+17)
    - 6\nu_2\left(9\nu_3^2 + 8\nu_3 - 17\right)
    \nonumber\\
    &\quad
    + 27\nu_3^3
    - 102\nu_3^2 + 105\nu_3-193]
  ,\\
  \mathcal{N}_{32}
  &=
  2
  - \frac{\varepsilon}{24} [ -3(9 \nu_3+2) \nu_2^2
    -\nu_2 \left(54 \nu_3^2+45\nu_3-62\right)
    -90 \nu_3^2+80 \nu_3-183
    \nonumber\\
    &\quad
    -2\nu_1(3\nu_2-6\nu_3+5)]
  +\frac{\varepsilon}{8}\nu_3 \left(-9 \nu_2 \nu_3+\nu_2+9 \nu_3^2-34 \nu_3
  +13\right),\\
  \mathcal{N}_{33}
  &=
  -\frac{\varepsilon}{8} \sqrt{3} \nu_2 [ \nu_2 (4-9 \nu_3)+25 \nu_3-8 ]
  +\frac{\varepsilon}{8} \sqrt{3}\nu_3 [\nu_2 (9 \nu_3-1)+9 \nu_3^2-18 \nu_3
    \nonumber\\
    &\quad
    +5]
  ,\\
  \mathcal{N}_{34}
  &=
  \frac{\varepsilon}{8} \sqrt{3} \nu_2 (9 \nu_2-13) (\nu_2+2 \nu_3-1)
  + \frac{\varepsilon}{8} \sqrt{3} \nu_3 (9 \nu_3-13) (2 \nu_2+\nu_3-1),\\
  \mathcal{N}_{35}
  &=
  \frac{3\varepsilon}{32} \sqrt{3} \nu_2 [ 15 \nu_2^2+\nu_2 (24 \nu_3-26)
    -3 \nu_3^2-34 \nu_3+16]
  \nonumber\\
  &\quad
  + \frac{3\varepsilon}{32} \sqrt{3} \nu_3[ -3 \nu_2^2+\nu_2 (24 \nu_3-34)
    +15 \nu_3^2-26\nu_3+16],\\
  \mathcal{N}_{36}
  &=
  \frac{3\sqrt{3}\nu_2}{4}
  - \frac{\varepsilon}{32} \sqrt{3} \nu_2 [ 36 \nu_2^2+ \nu_2 (21 \nu_3-4)
    +54 \nu_3^2+53 \nu_3+103
    \nonumber\\
    &\quad
    + \nu_1(3\nu_2-6\nu_3+5)]
  - \frac{3}{4}\sqrt{3}\nu_3
  + \frac{3\sqrt{3}\varepsilon}{32}\nu_3[17 \nu_2^2 + 4 \nu_2 (8 \nu_3-5)
    + 26 \nu_3^2
    \nonumber\\
    &\quad
    -28 \nu_3 + 52],\\
  \mathcal{N}_{37}
  &=
  \frac{9\nu_2}{4}
  - \frac{3\varepsilon}{32} \nu_2 [39\nu_2^2+\nu_2 (39\nu_3-64)+60 \nu_3^2
    -37 \nu_3+150
    \nonumber\\
    &\quad
    +\nu_1(3\nu_2-6\nu_3+5)]
  + \frac{9}{4}\nu_3
  - \frac{3\varepsilon}{32} \nu_3 [66\nu_2^2 + 6\nu_2(7 \nu_3-8)
    +36 \nu_3^2-66 \nu_3
    \nonumber\\
    &\quad
    +155],\\
  \mathcal{N}_{41}
  &=
  - 2-
  \frac{\varepsilon}{24}[ -6 \nu_2^2 (9 \nu_3+17)
    - 6\nu_2\left(9\nu_3^2 + 8\nu_3 - 17\right)
    + 27\nu_3^3 - 102\nu_3^2
    \nonumber\\
    &\quad
    + 105\nu_3-193]
  ,\\
  \mathcal{N}_{42}
  &=
  - \frac{\varepsilon}{8}\nu_3 \left(-9 \nu_2 \nu_3+\nu_2+9 \nu_3^2-34 \nu_3
  +13\right),\\
  \mathcal{N}_{43}
  &=
  - \frac{\varepsilon}{8} \sqrt{3}\nu_3 [\nu_2 (9 \nu_3-1)+9 \nu_3^2-18 \nu_3+5],
  \\
  \mathcal{N}_{44}
  &=
  - \frac{\varepsilon}{8} \sqrt{3} \nu_3 (9 \nu_3-13) (2 \nu_2+\nu_3-1),\\
  \mathcal{N}_{45}
  &=
  - \frac{3\varepsilon}{32} \sqrt{3} \nu_3[ -3 \nu_2^2+\nu_2 (24 \nu_3-34)
    +15 \nu_3^2-26\nu_3+16],\\
  \mathcal{N}_{46}
  &=
  \frac{3}{4}\sqrt{3}\nu_3
  -\frac{3\sqrt{3}\varepsilon}{32}\nu_3[17 \nu_2^2 + 4 \nu_2 (8 \nu_3-5)
    + 26 \nu_3^2-28 \nu_3 + 52],\\
  \mathcal{N}_{47}
  &=
  - \frac{9}{4}\nu_3
  + \frac{3\varepsilon}{32} \nu_3 [66\nu_2^2 + 6\nu_2(7 \nu_3-8)
    +36 \nu_3^2-66 \nu_3+155],
\end{align}
$\mathcal{N}_{51} = \mathcal{N}_{62} = \mathcal{N}_{73} = 1$, and the others
are 0.

\section{Relation between perturbations to size and orbital frequency}
\label{App:eigenv}

Equation \eqref{Eq:solPN} can be rewritten as
\begin{align}
  \bs{\chi}
  &=\mathrm{Q}E_1\mathrm{Q}^{-1}\bs{\chi}_0
  + e^{\omega_{\text{N}}t\lambda_{1+}}\mathrm{Q}E_2\mathrm{Q}^{-1}\bs{\chi}_0
  + e^{\omega_{\text{N}}t\lambda_{1-}}\mathrm{Q}E_3\mathrm{Q}^{-1}\bs{\chi}_0
  + e^{\omega_{\text{N}}t\lambda_{2+}}\mathrm{Q}E_4\mathrm{Q}^{-1}\bs{\chi}_0
  \nonumber\\
  &\quad
  + e^{\omega_{\text{N}}t\lambda_{2-}}\mathrm{Q}E_5\mathrm{Q}^{-1}\bs{\chi}_0
  + e^{\omega_{\text{N}}t\lambda_{3+}}\mathrm{Q}E_6\mathrm{Q}^{-1}\bs{\chi}_0
  + e^{\omega_{\text{N}}t\lambda_{3-}}\mathrm{Q}E_7\mathrm{Q}^{-1}\bs{\chi}_0,
  \label{eq:solution1}
\end{align}
where $E_i$ is a $7 \times 7$ matrix with the components as
\begin{align}
  E_i = \left\{
  \begin{array}{cl}
    1 & (\text{for the $ii$-component}) ,\\
    0 & (\text{for the others}) .
  \end{array}
  \right.
\end{align}
The regular matrix $Q$ is given by
\begin{align}
  Q=(\bs{v}\,\, \bs{v}_{1+}\,\, \bs{v}_{1-}\,\, \bs{v}_{2+}\,\, \bs{v}_{2-}\,\,
  \bs{v}_{3+}\,\, \bs{v}_{3-}),
\end{align}
where $\bs{v}_a$ are the eigenvectors corresponding to the eigenvalues
$\lambda_a$, and $\bs{v}$ is the eigenvector corresponding to the zero
eigenvalue.
Therefore, we have 
\begin{align}
  \left\{
  \begin{array}{l}
  \mathrm{Q}E_1
  =(\bs{v}\,\,\bs{0}\,\,\bs{0}\,\,\bs{0}\,\,\bs{0}\,\,\bs{0}\,\,\bs{0}),\\
  \mathrm{Q}E_2
  =(\bs{0}\,\,\bs{v}_{1+}\,\,\bs{0}\,\,\bs{0}\,\,\bs{0}\,\,\bs{0}\,\,\bs{0}),\\
  \mathrm{Q}E_3
  =(\bs{0}\,\,\bs{0}\,\,\bs{v}_{1-}\,\,\bs{0}\,\,\bs{0}\,\,\bs{0}\,\,\bs{0}),\\
  \mathrm{Q}E_4
  =(\bs{0}\,\,\bs{0}\,\,\bs{0}\,\,\bs{v}_{2+}\,\,\bs{0}\,\,\bs{0}\,\,\bs{0}),\\
  \mathrm{Q}E_5
  =(\bs{0}\,\,\bs{0}\,\,\bs{0}\,\,\bs{0}\,\,\bs{v}_{2-}\,\,\bs{0}\,\,\bs{0}),\\
  \mathrm{Q}E_6
  =(\bs{0}\,\,\bs{0}\,\,\bs{0}\,\,\bs{0}\,\,\bs{0}\,\,\bs{v}_{3+}\,\,\bs{0}),\\
  \mathrm{Q}E_7
  =(\bs{0}\,\,\bs{0}\,\,\bs{0}\,\,\bs{0}\,\,\bs{0}\,\,\bs{0}\,\,\bs{v}_{3-}),
  \end{array}
  \right.
\end{align}
and Eq.\eqref{eq:solution1} is calculated as
\begin{align}
  \bs{\chi}
  &= c_1\bs{v}
  + e^{\omega_{\text{N}}t\lambda_{1+}}c_2\bs{v}_{1+}
  + e^{\omega_{\text{N}}t\lambda_{1-}}c_3\bs{v}_{1-}
  \nonumber\\
  &\quad
  + e^{\omega_{\text{N}}t\lambda_{2+}}c_4\bs{v}_{2+}
  + e^{\omega_{\text{N}}t\lambda_{2-}}c_5\bs{v}_{2-}
  + e^{\omega_{\text{N}}t\lambda_{3+}}c_6\bs{v}_{3+}
  + e^{\omega_{\text{N}}t\lambda_{3-}}c_7\bs{v}_{3-},
\end{align}
where  $\bs{c}=(c_1, c_2, \cdots, c_7)\equiv \mathrm{Q}^{-1}\bs{\chi}_0$.
By setting the coefficients as 
\begin{align}
c_1\bs{v} &\equiv (\cdots, C_{41}, C_{11}, C_{21}, C_{31}) , \\
c_2\bs{v}_{1+} &\equiv (\cdots, C_{42}, C_{12}, C_{22}, C_{32}) , \\
c_3\bs{v}_{1-} &\equiv (\cdots, C_{43}, C_{13}, C_{23}, C_{33}) , \\
c_4\bs{v}_{2+} &\equiv (\cdots, C_{44}, C_{14}, C_{24}, C_{34}) , \\
c_5\bs{v}_{2-} &\equiv (\cdots, C_{45}, C_{15}, C_{25}, C_{35}) , \\ 
c_6\bs{v}_{3+} &\equiv (\cdots, C_{46}, C_{16}, C_{26}, C_{36}) , \\ 
c_7\bs{v}_{3-} &\equiv (\cdots, C_{47}, C_{17}, C_{27}, C_{37}) , 
\end{align}
we obtain the solution of Eq. \eqref{Eq:sol1PN}.

In order to consider $C_{41}$, which corresponds to the unique secular term, 
therefore, let us focus on the zero eigenvector $\bs{v}$.
By the definition of the zero eigenvector, 
we have the equation as
\begin{align}
\mathcal{N}\bs{v}=\bs{0} .
\end{align}
By the straightforward calculations, 
we obtain the equations for the components of $\bs{v}$ as 
\begin{align}
c_1 \bs{v} &= (0, 0, 0, C_{41}, C_{11}, C_{21}, C_{31}) , \\
C_{21} &= 0, \\
C_{31} &= 
- \frac{\sqrt{3}}{24} 
\left(-10\nu_1^2+5\nu_2^2+5\nu_3^2+4\nu_2\nu_3-2\nu_1\nu_2 - 2\nu_1\nu_3 \right) 
\varepsilon C_{11} , \\
C_{41} &= - \frac32 \left[ 1 - \frac{5}{48} ( 29 - 14 V ) \varepsilon \right] 
C_{11} .
\label{C41C11}
\end{align}
Therefore, $C_{41}$ is not independent of $C_{11}$, 
and hence 
the change of the size corresponding to $C_{11}$ leads to 
the change of the orbital frequency $C_{41}$ 
regarding the energy and angular momentum changes.

In fact, Eq. \eqref{C41C11} can be derived from 
Eqs. \eqref{Eq:ofreq1}-\eqref{Eq:ofreq3}.
Let us consider a perturbation $\ell \to \ell ( 1 + x )$ 
in Eq. \eqref{Eq:ofreq2}.
This leads to the perturbation to $\varepsilon$ as 
\begin{align}
\varepsilon \to \varepsilon ( 1 - x ) ,
\end{align}
at the leading order.
Therefore, 
the orbital frequency is perturbed as 
\begin{align}
\omega \to \omega_{\rm N} \left[ 1 + \tilde{\omega}_{\rm PN} 
- \frac32  \left( 1 + \frac{5}{3} \tilde{\omega}_{\rm PN} \right) x \right].
\end{align}
Replacing $x \to C_{11}$ in the last term, 
we obtain the perturbations to the orbital frequency 
equivalent to Eq. \eqref{C41C11}.



\end{document}